\RequirePackage{fix-cm}
\documentclass[smallextended]{svjour3}       
\smartqed  

\usepackage[dvipdfmx]{graphicx}
\usepackage{multirow}
\usepackage{here}
\usepackage{color}
\usepackage[normalem]{ulem}
\usepackage{amsmath}
\begin{document}

\title{Performances of a Compact Shielded Superconducting Magnet for Continuous Nuclear Demagnetization Refrigerator
}

\titlerunning{Performances of a Compact Shielded SC Magnet for CNDR}        

\author{S. Takimoto$^1$ \and
        R. Toda$^1$ \and
        S. Murakawa$^1$ \and
        Hiroshi Fukuyama$^{1, 2}$
}


\institute{S. Takimoto \at
              \email{takimoto@crc.u-tokyo.ac.jp}
              \and
              H. Fukuyama \at
              \email{hiroshi@phys.s.u-tokyo.ac.jp}
              \and
              $^1$ Cryogenic Research Center, The University of Tokyo, 2-11-16 Yayoi, Bunkyo-ku, Tokyo, Japan\\
              $^2$ Department of Physics, The University of Tokyo, 7-3-1 Hongo, Bunkyo-ku, Tokyo, Japan \\
}

\date{Received: date / Accepted: date}

\maketitle

\begin{abstract} 
We have successfully developed and tested a compact shielded superconducting (SSC) magnet with a FeCoV magnetic shield.
This was developed for the PrNi$_5$ based nuclear demagnetization refrigerator which can keep temperatures below 1 mK continuously (CNDR) [Toda $\it{et~al}$., J. Phys.: Conf. Ser. $\bf{969}$, 012093 (2018)]. 
The clear bore diameter, outer diameter, and total length of the SSC magnet are 22,  42 and 169~mm, respectively, and it produces the maximum field of 1.38~T at an electric current of 6~A.
In order to realize both the compactness and the high shielding performance, we carefully chose material and optimized design of the magnetic shield by numerical simulations of the field distribution based on measured magnetization curves of several candidate materials with high permeability.
We also measured a heat generated by sweeping the SSC magnet in vacuum to be 230~mJ per field cycle. 
This value agrees very well with an estimation from the measured magnetic hysteresis of the superconducting wire used to wind the magnet.

\end{abstract}

\section{Introduction}
\label{intro}
The sub-mK temperature environment is important for studies of unique quantum phases and phase transitions in liquid and solid $^3$He and other condensed matters. 
It is also useful even in broader fields, for example, those in which higher sensitivities in detection of X-ray or microwave are demanded, because thermal noises and heat capacities are extremely small at such temperatures. 
In order to realize much easier access to sub-mK temperatures for non-experts, we are developing a compact and continuous nuclear demagnetization refrigerator (CNDR)~\cite{Ref:Toda} using PrNi$_5$, a hyperfine enhanced nuclear magnet~\cite{Ref:Pobell}, as a coolant. 
CNDR has two independent PrNi$_5$ refrigeration units which are connected in series between the sample stage and the mixing chamber of dilution refrigerator through two superconducting heat switches.
It can keep a constant temperature below 1 mK continuously with a cooling power larger than 10~nW.

To take full advantage of the design concept of CNDR, it is crucial to develop a compact, low heat dissipation and high performance shielded superconducting (SSC) magnet.
Specifications required for the SSC magnet of our CNDR are the followings: (i) maximum magnetic field produced by a current of $I = 6$~A: $B_{\mathrm{max}} \geq$~1.2~T, (ii) clear bore diameter: $D_{\mathrm{bore}} = 22$~mm, outer diameter: $D_{\mathrm{od}} = 42$~mm, and total length: $L = 169$~mm, (iii) fringe fields at positions 50~mm away from the shield surfaces: $B \leq$~1~mT, and (iv) heat generation rate when the sweep rate is 1~mT/s: $\dot{Q} \leq$~1~mW.
In this article, we describe design details and test results of the SSC magnet which meets all the requirements listed above.
This magnet can be used not only for CNDRs in laboratory but also for other purposes where the size and weight of the magnet are severely restricted.

\section{Design of the SSC magnet}
\label{sec:1} 
To reduce $\dot{Q}$, it is essential to wind the solenoid with a superconducting (SC) wire consisting of smaller diameter filaments (see later discussions).
Also, to reduce the operation current, it is necessary to increase the coil constant by using a smaller overall diameter wire.
Among commercial SC wires, we chose the 0.14~mm diameter multifilamentary (54 filaments) NbTi wire with Cu clad (Cu:NbTi = 1.3:1) of Supercon, Inc (product No. 54S43).
It was wound on a copper bobbin up to 26 complete layers of 140~mm long.
The total length of the wire used is 2.1~km.
The inner and outer diameters of the winding part are 24 and 31~mm, respectively.

\begin{figure}[t]
\begin{center}
\includegraphics[width=100mm]{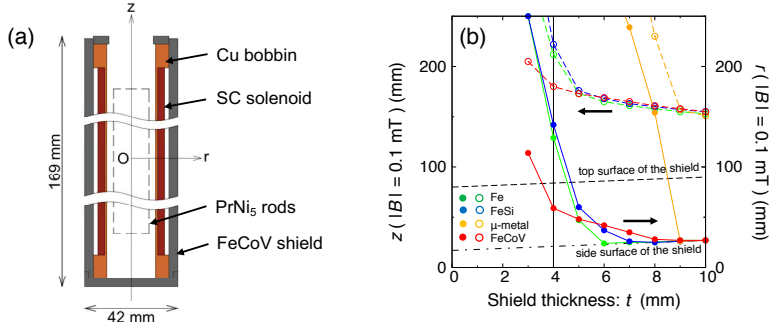}
\caption{(a) Cross section of the SSC magnet. (b) The $z$ ($r$) distances from the magnet center, beyond which the fringe field $|B|$ exceeds 0.1~mT, are shown by the open (closed) symbols and plotted as a function of shield thickness. The results are obtained by numerical simulations for four different shield materials, i.e., Fe, FeSi, $\mu$-metal, and FeCoV. The dashed and dotted lines represent locations of the top and side surfaces of the shield.}
\label{fig:cross section}
\end{center}
\end{figure}

To design the magnetic shield, we conducted numerical simulations of field distributions produced by solenoids wound by the above mentioned SC wire surrounded by two different types of cylindrical shields with the fixed dimensions ($D_{\mathrm{bore}}$, $D_{\mathrm{od}}$, and $L$) using the open source software of the finite element method (FEMM: Finite Element Method Magnetics by David Meeker). 
The results indicate that, when it is shielded by a cylinder made of a high permeability ($\mu$) and high saturation magnetization ($M_{\mathrm{sat}}$) material~\cite{Ref:Shirron}, the coil constant can be twice as much as that when shielded by an active shield coil~\cite{Ref:Israelsson,Ref:Milward}. 
We thus chose the former shielding type with two end caps as shown in Fig.~\ref{fig:cross section}(a).
Note that the upper end cap has a 20~mm diameter open hole through which thermal links for the PrNi$_5$ rods extend, which degrades the shielding performance considerably.

Next, in order to select a suitable high $\mu$ material and to optimize the shield thickness, we made numerical simulations of field distributions of SSC magnets with shields made of four candidate materials i.e., Fe, FeSi, $\mu$-metal, and FeCoV, of various thicknesses.
In the simulations, we used typical $\mu$ and $M_{\mathrm{sat}}$ values preinstalled in FEMM.
Fig.~\ref{fig:cross section}(b) shows simulation results for the axial ($z$) and radial ($r$) distances from the magnet center, beyond which the fringe field becomes less than 0.1~mT when $B_{\mathrm{max}} = 1.2$~T.
They are plotted as a function of the shield thickness $t$.
The dashed and dotted straight lines represent locations of the top and side surfaces of the shield.
Larger deviations from the lines for the simulation data mean higher fringe fields or incomplete shielding.
For all the materials, there is a critical thickness ($t_c$) below which such a deviation rapidly increases because of saturation of $M$ in the shield material.
To reduce the weight of magnet, it is better to use materials with lower $t_c$.
So we decided to shield the solenoid by a $t = 4$~mm thick FeCoV cylinder (Fe-49at\%, Co-49at\%, V-2at\%; Tohoku Steel Co., Ltd.), which meets all the requirements for our purpose.
The total weight of the SSC magnet we made is 1.2~kg.
Note that $t_c < 4$~mm for fringe fields at positions just outside of the closed bottom cap.

\section{Performances}
\label{sec:2}
\subsection{Magnetic field profiles}
\begin{figure}
\centering
\includegraphics[width=100mm]{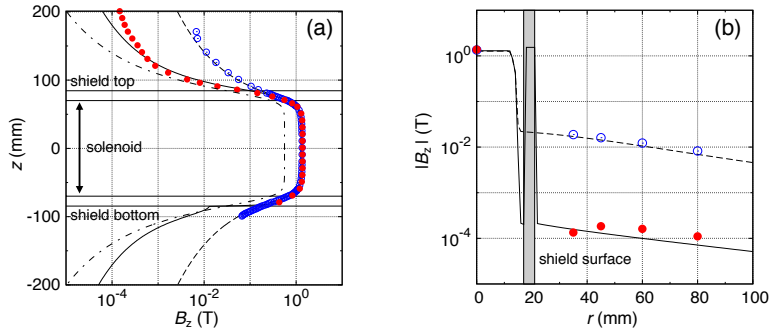}
\caption{(a) On-axis $B_z(z, r=0)$ and (b) radial $B_z(z=0, r)$ field profiles with (red closed circles) and without (blue open circles) the shield measured at $T = 4.2$~K. The solid (dashed) line represents the result of numerical calculation with (without) the shield. The dash-dot line in (a) is the calculated $B_z(z, r=0)$ profile for the magnet of the active shield type (see text). The shaded area in (b) represents the FeCoV shield location.}
\label{fig:magnetic field}
\end{figure}

We immersed the constructed SSC magnet in liquid $^4$He ($T = 4.2$~K) and measured its fringe fields with a Hall probe (OH002-2HR; Matsushita Electronics Industry Co., Ltd.).
Figure~\ref{fig:magnetic field}(a) and (b) show the measured on-axis and radial field profiles, respectively. 
The red closed and blue open circles represent the data with and without the FeCoV shield, respectively. 
The $B_{\mathrm{max}}$ value with shield ($= 1.38$~T when $I = 6$~A) is 2~\% higher than that without shield, while it is reduced by a factor of two in the case of the active shield type (the dash-dot line in Figure~\ref{fig:magnetic field}(a)).
At least at $B \geq 1$~mT, the data agree well with the numerical simulations indicated by the solid (with shield) and the dashed (without shield) lines for both the axial and radial directions. 
At $z \geq 125$~mm (or $B \leq 1$~mT), the data are consistently larger than the simulations. 
This is presumably due to the stray field which was accidentally produced by an extra winding of the lead wire.
The positions, beyond which the fringe field exceeds 1~mT, are 36~mm above the upper shield cap along the magnet center line and much closer than a few mm from the shield sidewall.
These results meet the requirement (iii).

\subsection{Estimation of heat generation from magnetic hysteresis}

\begin{figure}[t]
\centering
\includegraphics[width=100mm]{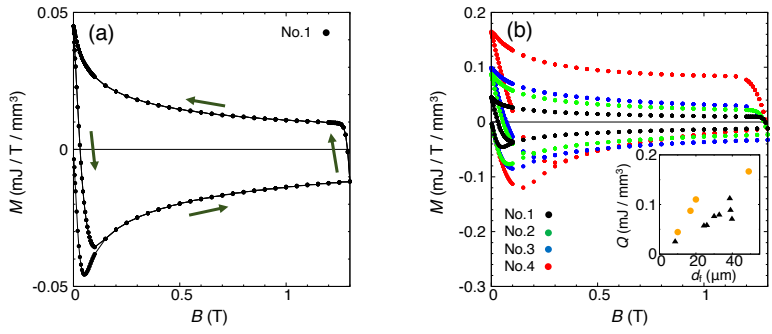}
\caption{Magnetization curve data of (a) the SC wire used in the SSC magnet (No.1 in Table~\ref{tab:SC wire}) and (b) of other types of wires (Nos.2-4 in Table~\ref{tab:SC wire}). The data were taken at $T =2$~K in vacuum. (inset) Q calculated from the magnetization curves shown in the main figure (orange closed circles) and Q from the previous research~\cite{Ref:SCwire} (black triangles).}
\label{fig:hysteresis}
\end{figure}

Magnetic hysteresises in the NbTi wire and the FeCoV shield are the main source of heat generation in the SSC magnet during sweep up and down.
The hysteresises originate from trapping of fluxoids in type II superconductors and from frictional motions of magnetic domain walls in ferromagnetic materials.
We measured the magnetization curve of a 6.5~mm long piece of the NbTi wire used in the present SSC magnet at $T =2$~K in vacuum using Magnetic Properties Measurement System (MPMS; Quantum Design, Inc.) (see Fig.~\ref{fig:hysteresis}(a)).
A heat $Q$ generated during one hysteresis loop ($B: 0 \to 1.2$~T~$\to 0$) can be calculated from

\begin{equation}
Q = \oint {MVdB},
\end{equation}

\noindent
where $M$ is the magnetization and $V$ is the volume of superconductor.
From this, we can estimate the heat generation of the whole solenoid per cycle to be $Q_{\mathrm{solenoid}} = 211 \pm2$~mJ, where 105~mJ is the heat generation during sweep up and 106~mJ is during sweep down.
We also estimated the heat generation of the whole FeCoV shield per cycle as $Q_{\mathrm{shield}} = 10 \pm1$~mJ by a similar procedure. 
Then the total heat generation of the SSC magnet per cycle is estimated to be $Q = Q_{\mathrm{solenoid}} + Q_{\mathrm{shield}} = 221$~mJ.
In the continuous operation mode of CNDR, the SSC magnet will be swept at a maximum rate of 1~mT/s.
Since the heat generation rate $\dot{Q}$ is proportional to $dM/dB$ under constant sweep rate, it becomes largest when $B$ returns to zero on sweeping down.
Otherwise, $\dot{Q}$ is roughly constant and would be of the order of several tens $\mu$W.
If this is the case, there is no problem to meet the requirement (iv).

\begin{table}[h]
\begin{center}
\caption{NbTi wires (Supercon Inc.) for which the magnetization curve data in Fig.~\ref{fig:hysteresis}(b) were taken. No.1 is the wire used in the present SSC magnet.}

\label{tab:SC wire}
\begin{tabular}{lcccc}
\hline\noalign{\smallskip}
& No.1 & No.2 & No.3 & No.4 \\
\noalign{\smallskip}\hline\hline\noalign{\smallskip}
Number of filaments & 54 & 18 & 54 & 1 \\
\noalign{\smallskip}
Filament diameter ($\mu$m) & 10 & 17 & 20 & 49 \\
\noalign{\smallskip}
Bare diameter (mm) & 0.114 & 0.114 & 0.229 & 0.079 \\
\noalign{\smallskip}
Insulator diameter (mm) & 0.140 & 0.140 & 0.254 & 0.102 \\
\noalign{\smallskip}
Clad material & Cu & CuNi & Cu & CuNi \\
\noalign{\smallskip}
Clad : NbTi ratio & 1.3 : 1 & 1.5 : 1 & 1.3 : 1 & 1.5 : 1 \\
\noalign{\smallskip}\hline
\end{tabular}
\end{center}
\end{table}

In Fig.~\ref{fig:hysteresis}(b), we show magnetization curves measured for four different types of SC wires listed in Table~\ref{tab:SC wire}.
It is known that $Q$ decreases with decreasing filament diameter~\cite{Ref:SCwire}.
This relation is also held in our data (the orange closed circles) but in a slightly different manner as shown in the inset.

\subsection{Direct measurement of heat generation rate}

\begin{figure}[b]
\includegraphics[width=\textwidth]{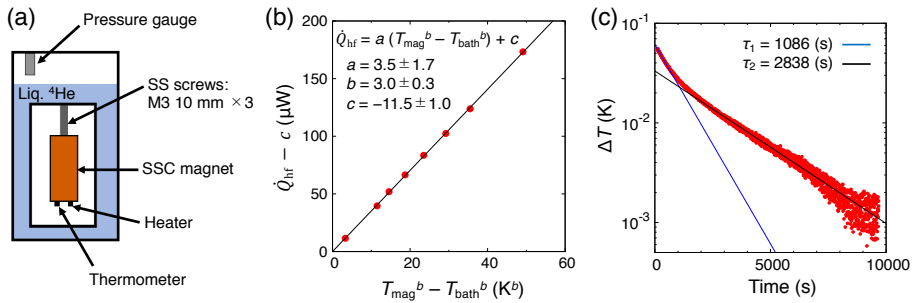}
\caption{(a) Experimental set up to measure the heat generation of the SSC magnet. (b) Measured relation among the heater power $Q$, the magnet temperature $T_{\mathrm{mag}}$, and the bath temperature $T_{\mathrm{bath}}$. The straight line is the fitting function (Eq.~\ref{eq:magnet temperature}) for the data. (c) Thermal relaxation process after a constant heat flow $\dot{Q}_{\mathrm{hf}} = 28$~$\mu$W was suddenly applied. The process involves two different time constants (see text).}
\label{fig:heater calibration}
\end{figure}

We have made direct measurements of $\dot{Q}$ generated by sweeping the SSC magnet at $T \approx 4$~K in vacuum.
The method is to monitor a temperature difference between the magnet and the thermal bath, which are connected with each other by a thermal link with a known thermal conductance.
The setup for the measurement is shown in Fig.~\ref{fig:heater calibration}(a), where the liquid helium bath is the thermal bath and the three 10~mm long M3 screws made of stainless steel (SS304) is the thermal link.
At the bottom of the shield end cap, a manganin heater (130~$\mathrm{\Omega}$) and a resistance thermometer (Cernox CX-1050-SD, Lake Shore Cryotronics, Inc.), which measures the temperature of the magnet ($T_{\mathrm{mag}}$), are attached.
The bath temperature ($T_{\mathrm{bath}}$) was determined by measuring the vapor pressure of $^4$He.

First of all, the thermal conductance $K$ of the SS screws was determined by measuring $T_{\mathrm{mag}}$ and $T_{\mathrm{bath}}$ at various $\it{constant}$ heat flows $\dot{Q}_{\mathrm{hf}}$.
Then the data were fitted to  

\begin{equation}
\dot{Q}_{\mathrm{hf}} = a(T_{\mathrm{mag}}^b - T_{\mathrm{bath}}^b) + c,
\label{eq:magnet temperature}
\end{equation}

\noindent
where $a = 3.5 \pm 1.7$~$\mu$WK$^{-b}$, $b = 3.0 \pm 0.3$, and $c = -11.5 \pm 1.0$~$\mu$W.
The fitting quality is sufficiently good as shown in Fig.~\ref{fig:heater calibration}(b).
Here, $c$ is an ambient heat leak to the liquid helium bath presumably through a remnant $^4$He exchange gas.
From this fitting, we determined $K (\mathrm{\mu W/K})$ as $K = abT^{(b-1)} = (10.5\pm 5.2) T^{2\pm0.3}$.
It is noted that this is slightly different from $K = (31.5\pm 2.2) T^{1.6\pm0.1}$ calculated from the previous report on the thermal conductivity of stainless steel 304~\cite{Ref:material}.

Fig.~\ref{fig:heater calibration}(c) shows a thermal relaxation process when a constant heat flow $\dot{Q}_{\mathrm{hf}} = 28$~$\mu$W was suddenly applied and then kept constant.
Apparently, the process involves two time constants, i.e., $\tau_1 = 1086 \pm10$~s and $\tau_2 = 2838 \pm22$~s.
$\tau_1$ would be related to internal thermalization within the SSC magnet.
$\tau_2$ should be associated with thermalization between the magnet and the thermal bath through the SS screws.
This is because the time constant calculated from an estimated heat capacity of the magnet and the known $K$ based on the two bath model agrees reasonably well with the measured $\tau_2$.
Note that the data shown in Fig.~\ref{fig:heater calibration}(b) were taken after waiting for much longer time than $\tau_2$.

\begin{figure}[t]
\centering
\includegraphics[width=60mm]{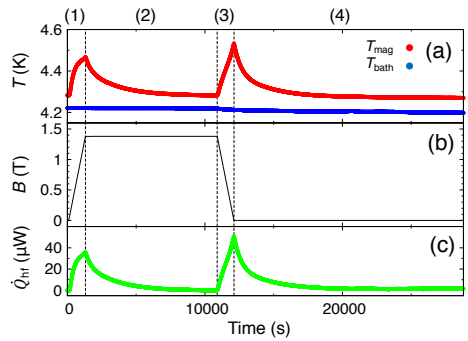}
\caption{(a) Time evolutions of $T_{\mathrm{mag}}$ and $T_{\mathrm{bath}}$ when the SSC magnet is swept in the sequence shown in (b). (c) Heat flow deduced from the data in (a) using Eq.~\ref{eq:magnet temperature}.}
\label{fig:magnet temperature}
\end{figure}

After knowing the relation among $\dot{Q}_{\mathrm{hf}}$, $T_{\mathrm{mag}}$ and $T_{\mathrm{bath}}$ (Eq.~\ref{eq:magnet temperature}), we measured time evolutions of $T_{\mathrm{mag}}$ and $T_{\mathrm{bath}}$ (see Fig.~\ref{fig:magnet temperature}(a)) without applying heater power this time but with sweeping the magnetic field $B$ of the SSC magnet in the following cycle (see Fig.~\ref{fig:magnet temperature}(b)): $B$ is (1) swept up from 0 to 1.38~T in 20~min, (2) kept constant at 1.38~T for the next 160~min until $T_{\mathrm{mag}}$ returns to the base temperature ($= 4.28$~K), (3) swept back to 0 in the next 20 min, and (4) kept constant at 0 in the next 280~min.
Figure~\ref{fig:magnet temperature}(c) shows an instantaneous heat flow $\dot{Q}_{\mathrm{hf}}$ calculated from $T_{\mathrm{mag}}$ and $T_{\mathrm{bath}}$ assuming Eq.~\ref{eq:magnet temperature} (quasi equilibrium assumption).
Reflecting the asymmetric hysteresis loop (Fig.~\ref{fig:hysteresis}(a)), $\dot{Q}_{\mathrm{hf}}$ is also asymmetric between the sweep up and down.

The cycle (1)$\sim$(4) was repeated three times successively.
By averaging heats integrated over processes (1) and (2) and those over (3) and (4), we obtain the total heat generation by the sweep up and down as $Q = 108 \pm 19$ and $118 \pm 11$~mJ, respectively.
These are in excellent agreement with those estimated from the magnetic hysteresis.
Such small heats can easily be absorbed by the still of dilution refrigerator through an annealed silver thermal link which should have a much higher thermal conductance than the SS screws by several orders of magnitude. 
This means that $T_{\mathrm{mag}}$ can be kept around $T= 0.8$~K.
Thus the radiation heat from the SSC magnet to the PrNi$_5$ nuclear stage should be approximately 0.1~nW, and this is much smaller than the expected cooling power ($\approx10$~nW) of CNDR at $T = 0.8$~mK~\cite{Ref:Toda}.

\section{Conclusions}
\label{sum}
We have successfully designed and constructed a high performance SSC magnet with a FeCoV magnetic shield, which is one of the key elements for development of the compact CNDR~\cite{Ref:Toda}.
It can produce $B_{\mathrm{max}} =$~1.38~T when $I = 6$~A with negligibly small fringe fields so that two SSC magnets can be located in close vicinity each other in CNDR. 
The measured heat generation due to sweeping the field at a rate of 1~mT/s is of the order of several tens $\mu$W, which is low enough to keep the magnet temperature at 0.8~K.



\begin{acknowledgements}
This work was financially supported by Grant-in-Aid for Challenging Exploratory Research (Grant No.~15K13398) from JSPS.
ST was supported by Japan Society for the Promotion of Science through Program for Leading Graduate Schools (MERIT).
\end{acknowledgements}

%
%



\end{document}